# Breaking the Baud Rate Ceiling of Electro-Optic Modulators Using Optical Equalization Technique

*Hengsong Yue, Nuo Chen, and Tao Chu\**

Hengsong Yue, Nuo Chen, and Tao Chu

College of Information Science and Electronic Engineering, Zhejiang University, Hangzhou 310027, China

E-mail: chutao@zju.edu.cn

This study presents an effective optical equalization technique for generating ultrahigh baud rate signals. The equalization technique was demonstrated using a dual-drive Mach-Zehnder modulator (DDMZM) with two-phase shifters having different bandwidths, which can be achieved by adjusting the structural design of the modulator or by incorporating varying bias voltages. When the two phase shifters are driven by appropriately sized driving signals, their partially offsetting response significantly reduces the rise and fall times of the output signal. The experiments were conducted using silicon DDMZMs without digital signal processing (DSP). In an experiment utilizing a low-speed silicon modulator, the on-off keying (OOK) modulating speed was improved from 30 to 90 Gbaud. In an experiment utilizing a high-speed silicon modulator, the OOK modulating speed was improved from 100 to 128 Gbaud, which approached the limit of our testing system. This remains the highest baud rate achieved by an all-silicon modulator without DSP. This technique breaks the baud rate ceiling of the modulator and has the potential to enable silicon modulators to operate at 200 Gbaud and beyond. The versatility of this method extends to a wide range of optoelectronic platforms, encompassing thin-film lithium niobate and indium phosphide modulators.

# 1. Introduction

Motivated by the exponential growth in data traffic, optical communications and interconnections have been extensively researched over the years owing to their high speed and economic feasibility.[1,2] Electro-optic modulators, which convert electrical signals into optical signals, are an essential component of these systems and play a crucial role in determining the communication speed.[3–5] Remarkable progress has been made in the development of modulators over the years.[6–18] These modulators can operate at frequencies up to one hundred GHz and reach data transmission rates up to several hundred Gbit/s by employing advanced modulation formats.[19–22] However, the limited baud rate of modulators has impeded further enhancement of the data transmission rate because of inherent bandwidth limitations, primarily caused by factors such as dielectric loss, conductor loss, parasitic resistance, and capacitance.[23–26] These parasitic components affect the signal rise and fall times, thereby limiting the baud rate.

Several methods have been proposed for addressing this issue. One method involves increasing the modulator bandwidth. For example, slow-wave electrodes have been used to achieve velocity matching between optical and electrical waves.[14,27] This is accomplished by using a periodic structure that slows the propagation of electrical signals in the electrode. In other works, substrate-removed modulators have been proposed, which employ an etching process to remove the silicon underneath the modulator, thereby enhancing the radio frequency performance.[16,28] Slow-light waveguides composed of Bragg gratings were incorporated into a silicon modulator, resulting in a record-high EO bandwidth and ultra-compact footprint.[29] However, these approaches have limitations, such as increased complexity, fabrication challenges, narrow optical bandwidth, potential loss issues, and weakened mechanical stability.

Another method is to use optical equalization techniques, which aim to compensate for the distortion introduced by the modulator and improve signal quality. For instance, optical-domain feed-forward equalization (FFE) using a segmented electrode Mach-Zehnder modulator (MZM) extends the MZM bandwidth without using equalized drivers.[30,31] The equalized traveling-wave MZM was also designed based on the concept of built-in feedback equalization, which makes use of the intrinsic delay of the traveling-wave electrodes and feedback circuit.[25] Additionally, an on-chip optical equalizer enables monolithic integration of a silicon in-phase/quadrature modulator with an optical equalizer, resulting in an open optical eye diagram and improved optical signal-to-noise ratio.[32] Moreover, the frequency response of the filter was mapped to the modulator response by embedding a finite-impulse response (FIR) filter in the slow-wave modulator, consequently achieving an enhanced electro-optical bandwidth.[33] However, the limitations of these methods include the dependence on segmented electrode Mach-Zehnder modulators, potential complexity and constraints in circuit design, reliance on integrated circuit fabrication, and the need for additional optical devices and circuit resources.

In this study, we propose an optical equalization technique to effectively overcome the inherent performance limitations of modulators, enabling the generation of ultrahigh baud rate signals. This innovation eliminated the need for resonant structures, thereby preserving a wide optical bandwidth. This technique uses a dual-drive Mach-Zehnder modulator (DDMZM) with two phase shifters of different bandwidths, which can be achieved through adjustments in the structural design or the incorporation of diverse bias voltages. A phase shifter with a high bandwidth is referred to as a high-speed phase shifter, and one with a low bandwidth is referred to as a low-speed phase shifter. The high-speed phase shifter was driven by a large amplitude signal, whereas the low-speed phase shifter was driven by a small amplitude signal. Their partially offsetting response reduced the rising and falling times of the DDMZM, showing the potential to significantly improve the quality of optical eye diagrams. Low-speed and high-speed modulator speed-up experiments were conducted using silicon DDMZMs, respectively, to investigate the characteristics and capabilities of this technique, as well as to demonstrate the generation of ultrahigh baud rate signals. In the low-speed modulator speed-up experiment, on-off keying (OOK) modulating speed was increased from 30 to 90

Gbaud. In the high-speed modulator speed-up experiment, owing to the limited bandwidth of the experimental system, the OOK modulating speed increased from 100 to 128 Gbaud. However, this still represents the highest baud rate attained using an all-silicon modulator without digital signal processing (DSP). Considering the notable progress made through experiments, it is feasible to allow a relatively low baud rate to increase to 200 Gbaud, or potentially even higher. The proposed optical equalization technique demonstrated the potential to achieve even higher baud rates for modulators, creating new possibilities for the development of ultrahigh speed optical communications.

## 2. Result and discussion

### 2.1. Theoretical Analysis.

**Figure 1** shows a schematic of the proposed optical equalization technique, which uses a DDMZM with a specialized configuration. Unlike the conventional DDMZM, this DDMZM was configured as a two-phase shifter with different bandwidths. Bandwidth differentiation can be accomplished through various methods, such as altering the length or doping concentration of the phase shifters and applying different junction bias voltages.[6,20,34,35] A high-speed phase shifter refers to one with high bandwidth, whereas a low-speed phase shifter refers to one with low bandwidth. The original electrical signal is amplified using two electrical amplifiers with distinct amplification factors, generating a pair of electrical signals with diverse amplitudes. An electrical signal with a higher amplitude was fed into the high-speed phase shifter, whereas an electrical signal with a lower amplitude was fed into the low-speed phase shifter. These electrical signals are loaded onto two optical carriers through phase shifters. The phase of the two optical carriers changes at different rates because the bandwidths of the phase shifters are different. Additionally, the depth of the phase shift varies owing to the different amplitudes of the electrical signals. In this context, the phase variation rate refers to the speed at which the phase of the optical carrier signal changes, whereas the phase shift depth represents the amount of phase shift applied to the optical carriers. Increasing the phase variation rate and decreasing the phase shift depth resulted in shorter rise and fall times. Assuming that the DDMZM operates at the quadrature point, the two optical signals interfere with each other on the multimode interference coupler, resulting in the subtraction of the phase shifts and their conversion into changes in the optical intensity. The phase-shift depth was suppressed during the subtraction process. The phase variation rate closely aligns with that of the high-speed phase shifter owing to the minor impact of the low-speed phase shifter. Consequently, there is a notable improvement in the rise and fall times compared with the high-speed phase shifter, resulting in the production of an equalized optical signal at the output.

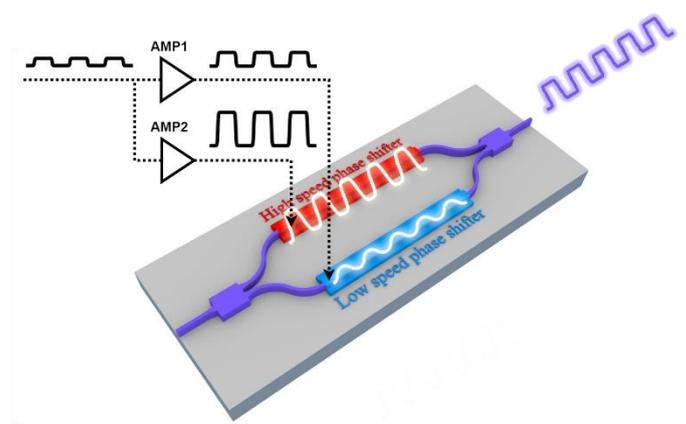

**Figure 1.** Schematic of the proposed optical equalization technique based on a DDMZM with specialized configuration.

The effectiveness of the optical equalization technique was demonstrated by simulating the time responses of electrical and optical signals, as illustrated in **Figure 2**. The 3 dB bandwidths of the high-speed and low-speed phase shifters were assumed as 60 GHz and 20 GHz, respectively. Their frequency response followed the silicon traveling wave MZM at velocity and impedance matching.[36] The two-phase shifters were driven by an ideal square-wave signal at 40 GHz with the normalized waveform depicted in Figure 2a. The time responses of the high-speed and low-speed phase shifters are shown in Figure 2b,c, respectively. They also depicted the intrinsic time response of the DDMZM without employing the aforementioned equalization technique. Benefit from higher bandwidth, the rise and fall times of the curve in Figure 2b are clearly shorter than the rise and fall times of the curve in Figure 2c. The amplitude of the curve in Figure 2c is notably suppressed because of the weak response of the low-speed phase shifter at high frequencies. Assuming that both the phase shifters have the same modulation efficiency and are driven by electrical signals with the same amplitude, the equalized response can be derived by subtracting Figure 2c from Figure 2b. This output is depicted in Figure 2d and was obtained using the multimode interference coupler shown in Figure 1. The equalized response displays a flat top, whereas Figure 2b,c exhibit consistently curved responses, indicating a remarkable enhancement in the signal shape compared with the inherent time response of the DDMZM without using the equalization technique.

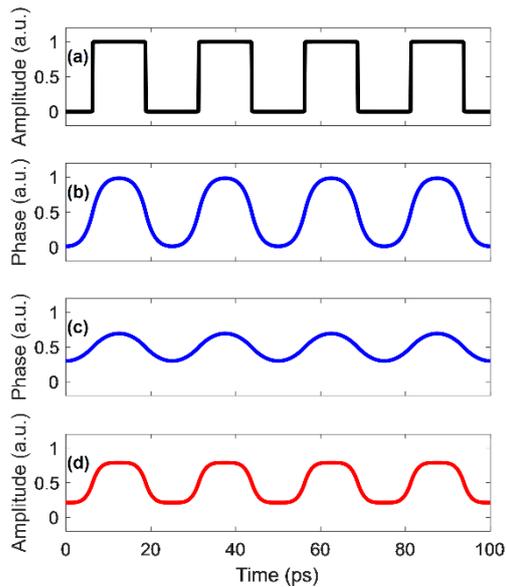

**Figure 2.** Time response of the a) electrical signal, b) high-speed phase shifter, c) low-speed phase shifter, and d) equalized output signal.

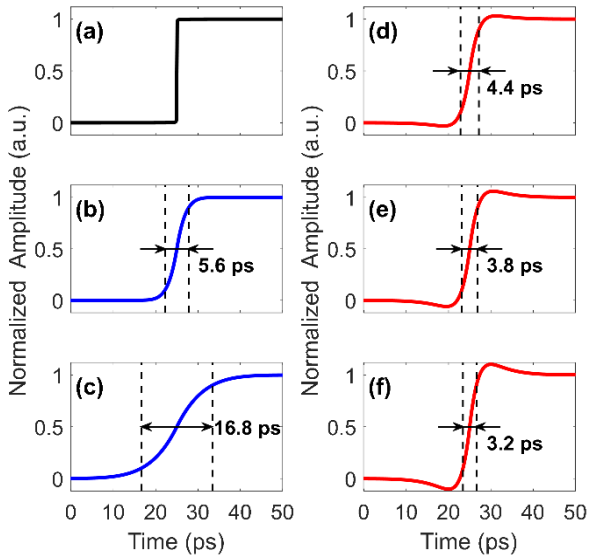

**Figure 3.** Simulated step response with and without equalization. a) Ideal step signal applied to the phase shifters. b,c) Step responses of the phase shifters at 3 dB bandwidth of 60 GHz and 20 GHz, respectively. d-f) Step responses of the equalized output signals at EF of 0.2, 0.3 and 0.4.

Furthermore, to evaluate the effectiveness of the optical equalization technique numerically, the step response was simulated, as shown in **Figure 3**. An ideal step signal is used to drive the phase shifters, and the normalized waveform is shown in Figure 3a. The resulting step responses of the high-speed and low-speed phase shifters at the 3 dB bandwidth of 60 GHz and 20 GHz are displayed in Figure 3b,c, respectively, revealing rise times of 5.6 and 16.8 ps. The equalization factor is defined as the ratio of the step-response amplitudes of the low-and high-speed phase shifters when the equalization technique is applied. It can be tuned by adjusting the proportion of the driving signal amplitudes for the low-speed and high-speed phase shifters. The step responses of the equalized output signals can be obtained by subtracting the response of the low-speed phase shifter multiplied by EF from the response of the high-speed phase shifter, owing to the interference of optical signals on the multimode interference coupler. The normalized equalization outcomes for EFs of 0.2, 0.3, and 0.4 are visually depicted in Figure 3d-f, with a clear overshoot on the waveform. The promising feature of signal overshoot has the potential to counteract performance degradation in other parts of the system, thereby enabling high-speed operation of the modulator. The corresponding rise times of 4.4, 3.8, and 3.2 ps correspond to bandwidths of 76, 88, and 105 GHz, respectively. This represents bandwidth enhancements of 27, 47, and 75% respectively, compared to the bandwidth of the high-speed phase shifter. This substantially enhanced effective bandwidth helps overcome the bandwidth limitation of the modulator, facilitating the generation of ultrahigh baud rate signals.

**2.2. Experimental Results.**

In this section, we demonstrate an optical equalization technique using silicon DDMZMs. The two identical phase shifters in the silicon DDMZM can easily be set up as high- and low-speed phase shifters because the bandwidth of the silicon electro-optic phase shifter is related to the bias voltage of the PN junction. This section presents two experiments: a low-speed modulator speed-up experiment to comprehensively investigate the characteristics and capabilities of the optical equalization technique, and a high-speed modulator speed-up experiment to explore the generation of ultrahigh baud-rate signals that could reach the bandwidth limit of the experimental system. The low-speed modulator speed-up experiment employs a low-speed silicon modulator with long phase shifters, whereas the high-speed modulator speed-up experiment uses a high-speed modulator with short phase shifters.

**2.2.1. Low-speed Modulator Speed-up Experiment.**

A low-speed silicon DDMZM was fabricated using the complementary metal-oxide-semiconductor process provided by the Semiconductor Manufacturing International Corporation. A micrograph of the silicon DDMZM is shown in **Figure 4**. The silicon DDMZM consists of two carrier-depletion-based phase shifters, one functioning at a high speed and the other at a low speed. Each phase shifter has a length of 3 mm. The Mach-Zehnder interferometer has a 50 μm path length difference between the two arms, which enables the adjustment of the optical carrier wavelength to change the operating point. The optical waveguide had a width of 450 nm, and a lateral PN junction was formed in the optical waveguide. The insertion loss of the device was 4 dB. Two grating couplers, each with an insertion loss of 4.1 dB, are used to couple the optical signal to and out of the chip.

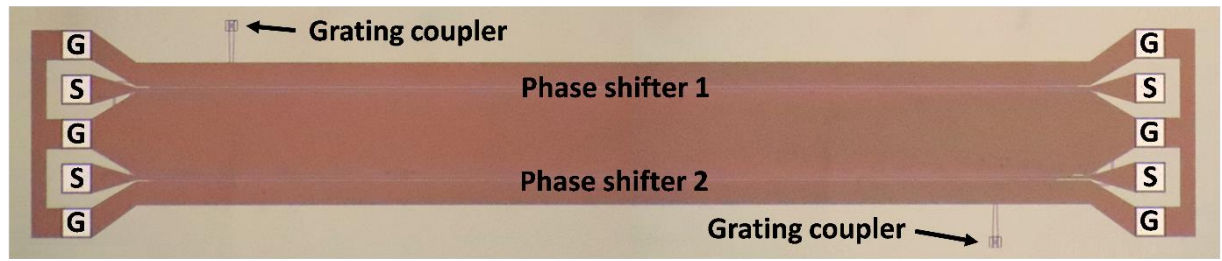

**Figure 4.** Micrograph of the low-speed silicon DDMZM. G: ground; S: signal.

**Figure 5** illustrates the experimental setup used to demonstrate the optical equalization technique on a low-speed silicon DDMZM. The optical carrier was generated by a C-band tunable laser source (Santec TSL550) and amplified with an erbium-doped fiber amplifier (EDFA) before being coupled to the silicon DDMZM via a grating coupler. Polarization alignment was achieved using a polarization controller. The 128 Gbps 2:1 multiplexer (SHF C603 B) accepted two serial data streams from the bit pattern generator (SHF 12104A), resulting in a high-baud-rate electrical OOK signal. The signal was amplified by a 60 GHz amplifier (SHF S804 B), and then split by a 40 GHz power splitter. One output was passed through an electrical attenuator to obtain EF related to its attenuation ratio of the electrical attenuator. Subsequently, the bias voltages generated by the two independent DC power sources are combined with the two outputs via the bias tees. These outputs are subsequently loaded onto the silicon DDMZM using a 40 GHz GSGSG probe and terminated with external 50 Ω loads. The silicon DDMZM applies optical equalization technology while converting electrical signals into optical signals. At the output of the DDMZM, the modulated signal undergoes amplification from an additional EDFA after which it is linked to a digital serial analyzer (DSA). Optically equalized OOK eye diagrams without DSP were obtained using DSA.

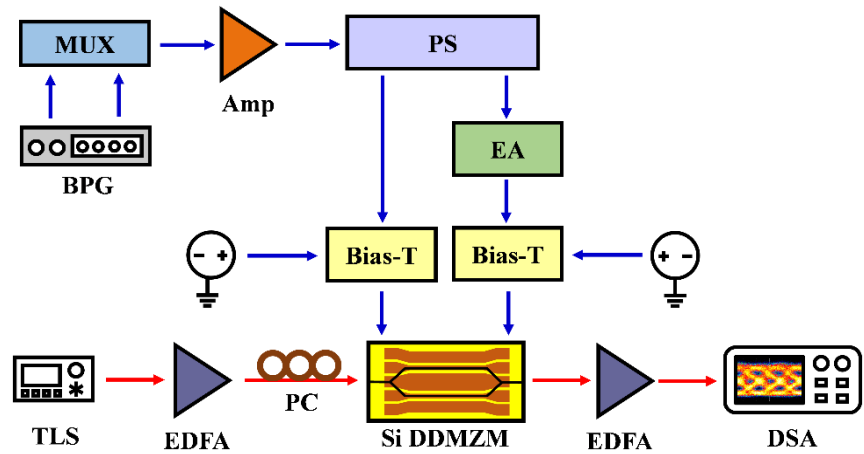

**Figure 5.** Experimental setup for demonstrating optical equalization technique on a low-speed silicon DDMZM. TLS: tunable laser source; EDFA: erbium-doped fiber amplifier; PC: polarization controller; DSA: digital serial analyzer; BPG: bit pattern generator; MUX: multiplexer; PS: power splitter; EA: electrical attenuator.

**Figure 6** shows the OOK eye diagrams obtained at various baud rates and EFs. In the EF = 0 scenario, which represents the un-equalized condition, only the high-speed phase shifter is driven, while the driving signal for the low-speed phase shifter is terminated with an external 50 Ω load following the power splitter. The high-speed phase shifter is driven with a Vpp of 3.6 V, whereas the Vpp values of the low-speed phase shifter are 1.3, 2.2, and 2.7 V at EFs of 0.36, 0.61, and 0.75, respectively. The bias voltage of the high-speed phase shifter was maintained at 6 V, whereas that of the low-speed phase shifter was continuously adjusted during the experiment to obtain an optimal eye diagram. This optimization is necessary because an excessively low bias voltage can cause a significant overshoot and degrade the eye diagram. Nonetheless, during the experiment, the bias voltage for the low-speed phase shifter was kept below 4 V. The maximum baud rate for the open-optical eye diagrams increased as EF increased, as shown in Figure 6. At EF = 0, the open-optical eye diagram had a maximum baud rate of only 30 Gbaud, which was limited by the bandwidth of the low-speed silicon DDMZM. Additionally, the diagram displays significant distortion and noise, indicating poor signal quality. However, when EF = 0.36, 0.61, and 0.75, the baud rates of the open-eye diagram are 60, 80, and 90 Gbaud, respectively. The increase in EF significantly improved the high-speed performance, which is in line with the bandwidth enhancement in the step-response simulation. In low-speed scenarios, theoretically, an increase in EF reduces the extinction ratio (ER), owing to the subtraction of the responses of the two-phase shifters. Indeed, when EF = 0, the limited bandwidth constrains the ER, leading to the gradual reduction in the ER observed only at EF = 0.36, 0.61, and 0.75. The experimental results indicated that the optical equalization technique approximately tripled the baud rate of the open-eye diagram, which can serve as a reference for its improvement ratio in high-speed scenarios.

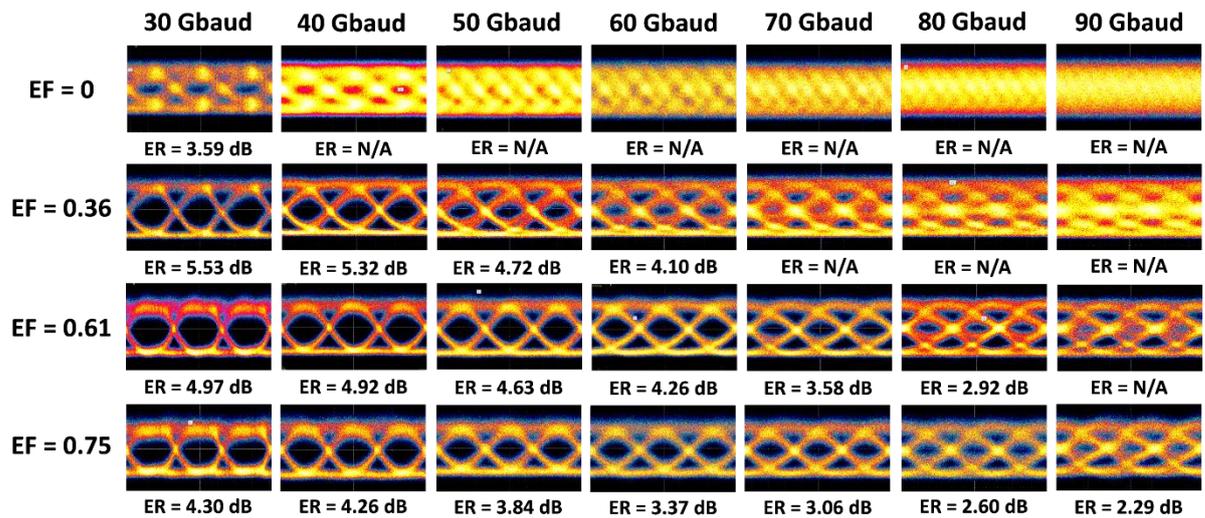

**Figure 6.** OOK eye diagrams at different baud rates and EFs for low-speed silicon DDMZM. ER: extinction ratio; N/A: not applied.

**2.2.2. High-speed Modulator Speed-up Experiment.**

**Figure 7** shows a micrograph of the high-speed silicon DDMZM fabricated on a silicon-on-insulator wafer at Advanced Micro Foundry. The DDMZM is an improved version of the previously introduced low-speed silicon DDMZM. Each phase shifter has a length of 1 mm to achieve a higher modulating speed. At the end of the traveling wave electrode, two 50 Ω on-chip terminators are connected in parallel to achieve a terminal impedance of 25 ohms. The silicon DDMZM exhibited an insertion loss of 3.1 dB, and each grating coupler had an insertion loss of 3.8 dB.

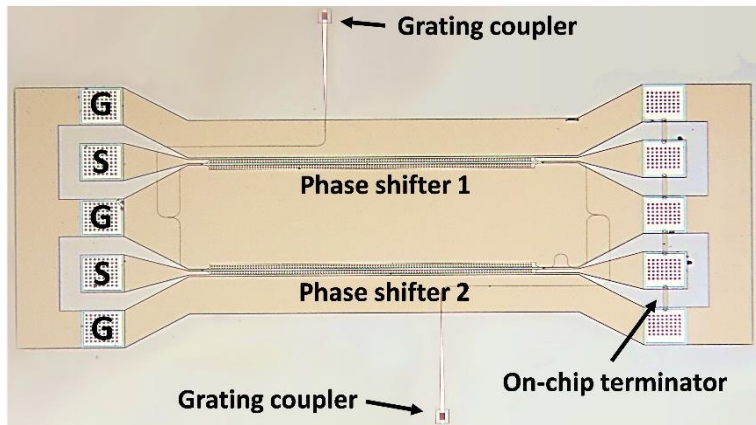

**Figure 7.** Micrograph of the high-speed silicon DDMZM. G: ground; S: signal.

The experimental setup shown in **Figure 8** is similar to that of the low-speed system described previously. They shared a C-band tunable laser source, two EDFAs, a polarization controller, a bit pattern generator, a multiplexer and a DSA. However, in this experiment, the pair of differential electrical OOK signals from the multiplexer were amplified using two 60 GHz amplifiers. One of the signals passed through a broadband balun (BAL-0067) to convert the phases of the two signals from opposite to identical and introduced an attenuation in the amplitude of the electrical signal, which served as an EF of 0.5. The resulting signals were then combined with the bias voltages via bias tees before being loaded onto a high-speed silicon DDMZM.

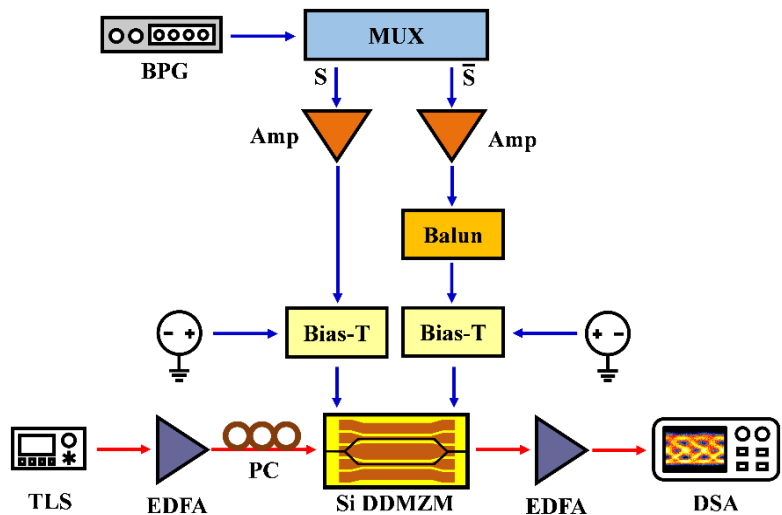

**Figure 8.** Experimental setup for demonstrating optical equalization technique on a high-speed silicon DDMZM. TLS: tunable laser source; EDFA: erbium-doped fiber amplifier; PC: polarization controller; DSA: digital serial analyzer; BPG: bit pattern generator; MUX: multiplexer.

**Figure 9**a,b show the OOK eye diagrams of the high-speed silicon DDMZM at 2 EFs of 0 and 0.5, respectively. The driving signal of the high-speed phase shifter has a Vpp of 5 V, whereas that of the low-speed phase shifter is 2.5 V when EF = 0.5. The bias voltages for the high-speed and low-speed phase shifters were 5 V and 3.2 V, respectively. When EF = 0, the maximum baud rate of the open-eye diagram is 100 Gbaud. The eye diagrams are significantly improved in Figure 9b, where the eye diagrams between 90 and 110 Gbaud are clearly open without significant degradation. Additionally, an open-eye diagram of 128 Gbaud was achieved with an ER of 1.89 dB, which was even higher than that of 90 Gbaud when EF = 0. The apparent degradation of the eye diagrams from 110 to 128 Gbaud can be attributed to the limited bandwidth of the electrical amplifiers and the 128 Gbps multiplexer, which restricts the maximum achievable baud rate to 128 Gbaud. Nevertheless, this is still the highest baud rate achieved without DSP for an all-silicon modulator. By referencing the efficacy of the optical equalization technique observed in the experiments, the implementation of this technique has the potential to enable silicon modulators to achieve baud rates of 200 Gbaud and beyond without requiring specifically designed modulators or equalizers.

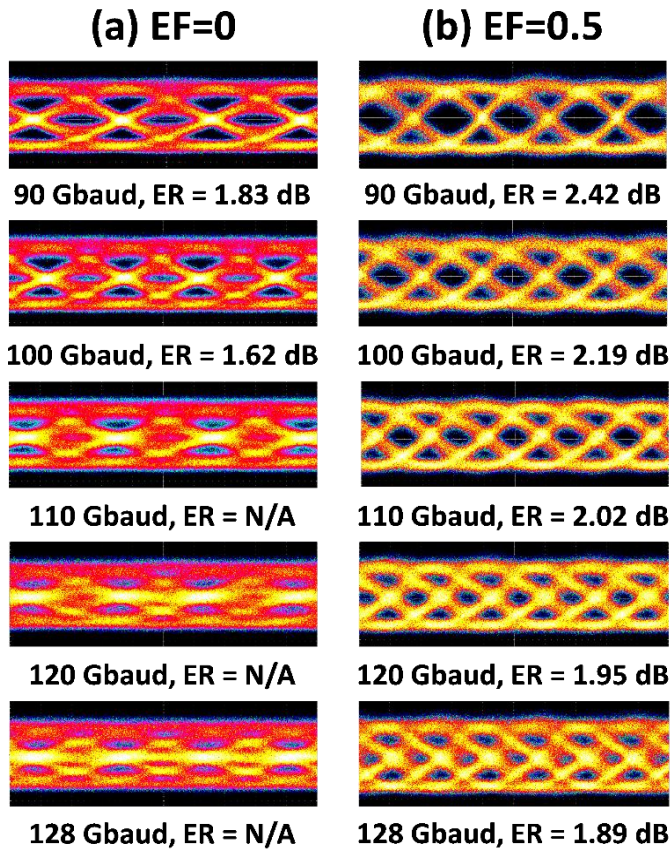

**Figure 9.** OOK eye diagrams at 2 EFs of a) 0 and b) 0.5 for high-speed silicon DDMZM.

## 3. Conclusions

This study proposes an optical equalization technique to overcome the performance limitations of modulators and generate signals with an ultrahigh baud rate. Optical equalization is achieved by making the two arms of the DDMZM have different bandwidths and then applying drive signals with appropriate amplitudes. The subtraction of the responses of the two arms considerably reduces the rise and fall times of the output signal, thereby significantly improving the signal quality. Both low-speed and high-speed modulator speed-up experiments were conducted using low-speed and high-speed silicon DDMZM, respectively. A low-speed modulator speed-up experiment demonstrated a substantial improvement ratio, increasing the baud rate of the generated signal from 30 to 90 Gbaud. However, in the high-speed modulator speed-up experiment, owing to the limited bandwidth of the experimental system, the baud rate of the generated signal only increased from 100 to 128 Gbaud. Nonetheless, this is the highest baud rate achieved using an all-silicon modulator without DSP. Referring to the improvements observed in the experiments, there is the potential to enhance the baud rate of silicon modulators to exceed 200 Gbaud. It should be noted that the optical equalization technique is not limited to silicon platforms, being highly versatile and adaptable to other integrated platforms and discrete devices. The breakthrough in the modulator baud rate paves the way for the applicability of silicon modulators in future ultrahigh speed communications and fosters advancements across various optoelectronic platforms. This represents a significant advancement in the development of future-oriented communication.

## 4. Experimental Section/Methods

*4.1. Sample fabrication.*

The low-speed silicon DDMZM was manufactured by Semiconductor Manufacturing International Corporation. The silicon modulator is comprised of two identical carrier-depletion phase shifters, each with a length of 3 mm and a fill factor of 91%. The intrinsic region was periodically inserted to prevent current from flowing through the optical waveguide. The traveling wave electrode has a signal line width of 10 μm and a 5 μm gap between the signal and ground lines. The lateral PN junction in the optical waveguide is offset by 50 nm from the center to the N-doped region. Heavily doped regions were used to form an ohmic contact with the metal electrode, spaced 750 nm from the edge of the optical waveguide, to obtain a tradeoff between optical absorption loss and modulating speed. The insertion loss of the low-speed silicon DDMZM was 4 dB.

The high-speed silicon DDMZM was manufactured by Advanced Micro Foundry, with each phase shifter having a length of 1 mm. The high-speed silicon DDMZM has a traveling wave electrode size equivalent to that of the low-speed version, with two on-chip terminators of 50 Ω placed in parallel at the end of the traveling wave electrode, resulting in a terminal impedance of 25 Ω. The optical waveguide had a width of 500 nm and slab thickness of 70 nm. Lateral PN junctions were formed at the center of the optical waveguide, whereas heavily doped regions were located 700 nm from the edge of the waveguide. The insertion loss of the high-speed silicon DDMZM was 3.1 dB.

**Acknowledgements**

The authors acknowledge the funding support from the National Natural Science Foundation of China (62035001).

**Conflict of Interest**

The authors declare no competing financial interests.

**Data Availability Statement**

The data that support the findings of this study are available from the corresponding author upon reasonable request.